\documentclass[aps,pra,twocolumn,showpacs]{revtex4}
\usepackage{bbm,amsmath,amssymb,amsbsy,graphicx,times,subfigure}

\usepackage[latin1]{inputenc}
\newcommand{\R}{\mathbbm{R}}

\newcommand{\sy}[1]{Sp_{(#1,\R)}}

\renewcommand{\det}{{\rm Det}\,}
\newcommand{\gr}[1]{\boldsymbol{#1}}
\newcommand{\be}{\begin{equation}}
\newcommand{\ee}{\end{equation}}
\newcommand{\bea}{\begin{eqnarray}}
\newcommand{\eea}{\end{eqnarray}}
\newcommand{\ket}[1]{|#1\rangle}
\newcommand{\bra}[1]{\langle#1|}

\newcommand{\N}{{\cal N}}
\newcommand{\D}{\Delta}
\newcommand{\sig}{\gr{\sigma}}
\newcommand{\bet}{\gr{\beta}}
\newcommand{\alp}{\gr{\alpha}}

\newcommand{\eq}[1]{Eq.~(\ref{#1})}

\newcommand{\ie}{\emph{i.e.}~}

\begin{document}
\title{Unitarily localizable entanglement of Gaussian states}
\date{January 10, 2005}
\author{Alessio Serafini}
\author{Gerardo Adesso}
\author{Fabrizio Illuminati}
\affiliation{Dipartimento di Fisica ``E. R. Caianiello'',
Universit\`a di Salerno, INFM UdR di Salerno, INFN Sezione di Napoli,
Gruppo Collegato di Salerno,
Via S. Allende, 84081 Baronissi (SA), Italy}


\begin{abstract}
We consider generic $m\times n$-mode bipartitions of continuous variable systems,
and study the associated bisymmetric multimode Gaussian states.
They are defined as $(m+n)$-mode Gaussian states invariant
under local mode permutations on the $m$-mode and $n$-mode
subsystems. We prove that such states are equivalent, under local
unitary transformations, to the tensor product of a two-mode state and of
$m+n-2$ uncorrelated single-mode states.
The entanglement between the $m$-mode and the $n$-mode blocks can then be
completely concentrated on a single pair of modes by means of
local unitary operations alone.
This result allows to prove that the PPT (positivity of the partial transpose)
condition is necessary and sufficient for the separability of
$(m + n)$-mode bisymmetric Gaussian states. We
determine exactly their negativity and identify a subset
of bisymmetric states whose multimode entanglement of formation can be
computed analytically. We consider explicit examples of pure and mixed
bisymmetric states and study their entanglement scaling with the
number of modes.
\end{abstract}
\maketitle

\section{Introduction}

In quantum information and computation science,
it is of particular relevance to provide
theoretical methods to determine the entanglement of systems
susceptible to encompass many parties.
Such an interest does not stem only from pure intellectual curiosity,
but also from practical needs in the
implementations of realistic information protocols.
This is especially true as soon as
one needs to encode two-party information in a multipartite structure in
order to minimize possible errors and decoherence effects
\cite{chuangnielsen,heiss}. The study of the structure of
multipartite entanglement poses many formidable challenges,
concerning both its qualification and quantification, and
so far little progress has been achieved for multi-qubit systems
and in general for multi-party systems in finite dimensional
Hilbert spaces. However, the situation looks somehow more promising
in the arena of continuous variable systems, where some aspects of
genuine multipartite entanglement can be at least qualitatively understood
in the study of the entanglement of multimode bipartitions.

In the present work we study in detail the entanglement properties of
multimode Gaussian states of continuous variable (CV) systems
(for an introduction to CV quantum information see Ref.~\cite{braunsteinreview}).
After the seminal analysis on the separability
of two-mode Gaussian states \cite{simon00,duan00},
much progress has been accomplished
on the separability conditions of multimode
Gaussian states under various bipartitions
\cite{werner01,giedkeprl01,giedkepra01,vloock03}.
On the other hand, much less is known on the
quantification of the entanglement of multimode, multipartite
Gaussian states \cite{vloock02}. In a previous
work \cite{adesso04}, we have presented a theoretical scheme to
exactly determine the entanglement of, pure or mixed
$(n + 1)$-mode Gaussian states,
under $1 \times n$-mode bipartitions, endowed with full or partial symmetries
under mode exchange. More recently, a measure of genuine multipartite
CV entanglement has been proposed \cite{contangle}, that extends
the approach introduced by Coffman, Kundu, and Wootters for multiqubit systems
\cite{valerie}, and possesses a precise operational 
meaning related to the optimal fidelity of teleportation in a continuous-variable
teleportation network \cite{adessoteleportation}.

In this paper we generalize the analysis introduced in Ref.~\cite{adesso04}
to bisymmetric $(m+n)$-mode Gaussian states of $m \times n$-mode bipartitions.
The main result of the present paper is
that the bipartite entanglement of bisymmetric $(m + n)$-mode Gaussian
states is {\em unitarily localizable}, {\em i.e.}~that, through local unitary operations,
it may be fully concentrated in a single pair
of modes, each of them owned by one of the two parties (blocks). Here the
notion of localizable entanglement is different from that
introduced by Verstraete, Popp, and Cirac for spin systems
\cite{vpclocal}. There, it was defined as the maximal entanglement
concentrable on two chosen spins through local {\em measurements} on all
the other spins. Here, the local operations that concentrate
all the multimode entanglement on two modes are {\em unitary} and
involve the two chosen modes as well, as parts of the respective blocks.

The consequences of the unitary localizability are manifold.
In particular, the PPT (positivity of the partial transpose)
criterion is proved to be a necessary and sufficient condition
for the separability of $(m + n)$-mode bisymmetric Gaussian states.
Moreover, the block entanglement
({\ie} the entanglement between blocks of modes) of
bisymmetric (generally mixed) Gaussian states can be determined.
The entanglement can be quantified by the logarithmic negativity in
the general instance because the PPT criterion holds, but
we will also show some explicit cases in which the
entanglement of formation between $m$-mode and $n$-mode parties
can be exactly computed.

The plan of the paper is as follows. In Sec.~\ref{gs} we introduce
the notation and review some basic facts about Gaussian states and
their entanglement properties. In Sec.~\ref{sfbisym} we show that a bisymmetric
Gaussian state reduces to the tensor product of a correlated
two-mode state and of uncorrelated single-mode states. In
Sec.~\ref{block} we exploit such a result to explicitly determine
the entanglement of bisymmetric Gaussian states. In Sec.~\ref{examp} the
scaling of the block entanglement and the evaluation of the
unitarily localizable entanglement involving different partitions of
(generally mixed) symmetric states are studied in detail.
Finally, in Sec.~\ref{conclu} we present some conclusions and
miscellaneous comments.

\section{Gaussian states of bosonic systems\label{gs}}

Let us consider a CV system, {\em i.e.~}a system described by an
infinite dimensional Hilbert space ${\cal H}=\bigotimes_{i=1}^{n}
{\cal H}_{i}$ resulting from the tensor product
of infinite dimensional Fock spaces ${\cal H}_{i}$'s.
Let $a_{i}$ be the annihilation operator acting on ${\cal H}_{i}$,
and $\hat x_{i}=(a_{i}+a^{\dag}_{i})$ and
$\hat p_{i}=(a_{i}-a^{\dag}_{i})/i$ be
the related quadrature phase operators.
The corresponding phase space variables will be denoted by $x_{i}$ and $p_{i}$.
Let us group together the operators $\hat x_{i}$ and $\hat p_{i}$ in a
vector of operators $\hat X = (\hat x_{1},\hat p_{1},\ldots,\hat x_{n},\hat p_{n})$.
The canonical
commutation relations (CCR) for the $\hat X_{i}$'s are encoded in
the symplectic form $\gr{\Omega}$
\[
[\hat X_{i},\hat X_j]=2i\Omega_{ij} \; ,
\]
\be
{\rm with}\quad\boldsymbol{\Omega}\equiv
\gr{\omega}^{\oplus n}
\; , \quad \boldsymbol{\omega}\equiv \left( \begin{array}{cc}
0&1\\
-1&0
\end{array}\right) \label{symform}\; .
\ee A complete description of a CV quantum state $\varrho$ can be
provided in terms of its symmetrically ordered characteristic
function $\chi$. If we define the displacement operator
$D_{\xi}=\exp(i\hat{X}^{\sf T} \gr{\Omega}\xi)$, with $\xi\in
\R^{2n}$, then the characteristic function $\chi$ associated to
$\varrho$ is given by $\chi(\xi)=\,{\rm Tr}[\varrho D_{\xi}]$. The
set of Gaussian states is, by definition, the set of states with
Gaussian characteristic functions. Therefore a Gaussian state
$\varrho$ is completely characterized by its first and second
statistical moments which form, respectively, the vector of first
moments $\bar X\equiv\left(\langle\hat X_{1} \rangle,\langle\hat
X_{1}\rangle,\ldots,\langle\hat X_{n}\rangle, \langle\hat
X_{n}\rangle\right)$ and the covariance matrix (CM) $\sig$ of
elements
\begin{equation}
\sigma_{ij}\equiv\frac{1}{2}\langle \hat{X}_i \hat{X}_j +
\hat{X}_j \hat{X}_i \rangle -
\langle \hat{X}_i \rangle \langle \hat{X}_j \rangle \, , \label{covariance}
\end{equation}
where, for any observable $\hat{o}$,
$\langle\hat o\rangle\equiv\,{\rm Tr}(\varrho\hat o)$.
First statistical moments can be arbitrarily adjusted by
local unitary operations, which do not affect any property related
to correlations or entropies. Therefore they will be unimportant
to our aims and we will set them to $0$ in the following, without
any loss of generality. Throughout the paper, $\gr{\sigma}$ will
stand for the covariance matrix of the Gaussian state
$\varrho$.\par

The positivity of $\varrho$ and the CCR entail the
following relation on the CM $\gr{\sigma}$ of a
quantum state $\varrho$
(``Robertson-Schr\"odinger'' uncertainty relation)
\begin{equation}
\boldsymbol{\sigma}+ i\boldsymbol{\Omega}\ge 0 \; ,
\label{bonfide}
\end{equation}
Inequality (\ref{bonfide}) is the
necessary and sufficient constraint $\boldsymbol{\sigma}$
has to fulfill to
be a {\em bona fide} CM \cite{simon87}. We mention
that such a constraint implies $\gr{\sigma}\ge0$.

The class of unitary transformations generated by second order
polynomials in the field operators (`second-order' operations)
is especially relevant in
manipulating Gaussian states. For a $n$-mode systems, such operators
may be mapped, through the so called {\em `metaplectic'}
representation, into the real symplectic group $\sy{2n}$
\cite{folland}, made up by linear operations acting on a linear
space (called `phase space' in analogy with classical Hamiltonian
dynamics), which preserves the symplectic form $\gr{\Omega}$ under
congruence:
\[
S\in \sy{2n}  \Leftrightarrow S^{\sf T} \gr{\Omega} S = \gr{\Omega}
\; .
\]
Symplectic operations preserve the Gaussian character of the input
state, acting linearly on first moments and by congruence on second
moments:
\[
\gr{\sigma} \rightarrow S^{\sf T} \gr{\sigma} S \; .
\]
Ideal squeezers and beam splitters are examples of
(respectively, `active' and `passive') symplectic transformations.

A tensor product of Hilbert spaces (and of `second-order' unitary operations)
is mapped into a direct sum of phase spaces (and of symplectic transformations).
Under a $m\times n$ mode partition, resulting from the direct sum of
phase spaces $\Gamma_1$ and $\Gamma_2$ with dimensions $2m$ and $2n$
respectively,
we will refer to a transformation $S_{l} = S_{1}\oplus S_2$, with
$S_{1} \in Sp_{(2m,\mathbb R)}$ and $S_{2} \in Sp_{(2n,\mathbb R)}$ acting on
$\Gamma_{1}$ and $\Gamma_{2}$, as to a ``local symplectic operation''.
The corresponding unitary transformation is
the ``local unitary transformation'' $U_{l}=
U_{1}\otimes U_2$.

Let us recall that, due to a theorem by Williamson \cite{williamson36},
the CM of a $n$--mode Gaussian state can always be
written as \cite{simon87}
\begin{equation}
\gr{\sigma}=S^{\sf T} \gr{\nu} S \; , \label{willia}
\end{equation}
where $S\in Sp_{(2n,\mathbb{R})}$ and $\gr{\nu}$ is the CM
\begin{equation}
\gr{\nu}=\,{\rm diag}({\nu}_{1},{\nu}_{1},\ldots,{\nu}_{n},{\nu}_{n}) \, ,
\label{therma}
\end{equation}
corresponding to a tensor product of thermal states with diagonal
density matrix $\varrho^{_\otimes}$ given by
\[
\varrho^{_\otimes}=\bigotimes_{i}
\frac{2}{\nu_{i}+1}\sum_{k=0}^{\infty}\left(
\frac{\nu_{i}-1}{\nu_{i}+1}\right)^k \ket{k}_{i}{}_{i}\bra{k}\; ,
\]
$\ket{k}_i$ being the $k$-th number state of the
Fock space ${\cal H}_{i}$. The dual (Hilbert space) formulation
of \eq{willia} then reads: $\varrho=U^{\dag}\,\varrho^{_\otimes}\,
U$, for
some unitary $U$.
The quantities $\nu_{i}$'s form the symplectic spectrum of
the covariance matrix $\gr{\sigma}$ and can be
computed as the eigenvalues of the matrix $|i\gr{\Omega}\gr{\sigma}|$
\cite{absolute}.
Such eigenvalues are in fact invariant under the action
of symplectic transformations on the matrix $\gr{\sigma}$.\\
The symplectic eigenvalues $\nu_{i}$ encode essential informations
on the Gaussian state $\varrho$ and provide powerful, simple
ways to express its fundamental properties. For instance,
provided that the CM $\sig$ satisfies $\sig\ge 0$, then
\[ {\nu}_{i}\ge1 \;
\]
is equivalent to the uncertainty relation (\ref{bonfide}).
We remark that the full saturation of the uncertainty principle can
only be achieved by pure $n$-mode Gaussian states, for which
$\nu_i=1\,\,\forall i=1,\ldots, n$. Instead, mixed states such that
$\nu_{i\le k}=1$ and $\nu_{i>k}>1$, with $1\le k\le n$, only
partially saturate the uncertainty principle, with partial
saturation becoming weaker with decreasing $k$. Such states are
minimum uncertainty mixed Gaussian states in the sense that the
phase quadrature operators of the first $k$ modes satisfy the
Heisenberg minimal uncertainty, while for the remaining $n-k$ modes
the state indeed contains some additional thermal and/or
Schr\"odinger--like correlations which are responsible for the
global mixedness of the state.

The symplectic eigenvalues are
clearly invariant under symplectic operations. Yet, it is often
advantageous to introduce other symplectic invariants,
which can be easily handled in terms of second statistical moments. In the
present work, dealing with a $n$-mode Gaussian state with CM $\sig$,
we will make use of the obvious invariant ${\rm Det}\,\sig$ (whose
invariance is a consequence of the fact that ${\rm Det}\,S=1$
$\forall S\in \sy{2n}$) and of
$\Delta_{\sig}=\sum_{i,j=1}^{n}\,{\rm Det}\,\sig_{ij}$, where the
$\sig_{ij}$ are $2\times 2$ submatrices of $\sig$: \be
\sig=\left(\begin{array}{ccc}
\sig_{11}&\cdot&\sig_{1n}\\
\vdots&\ddots&\vdots\\
\sig_{n1}&\cdots&\sig_{nn}
\end{array}\right) \label{subma} \; .
\ee The invariance of $\Delta_{\sig}$ in the multimode case follows
from its invariance in the case of two-mode states, proved in Ref.~\cite{serafozzi},
and from the fact that any symplectic transformation can be
decomposed as the product of two-mode transformations
\cite{agarwal}. The symplectic eigenvalues $\nu^{\mp}$ of a two-mode
Gaussian state are simply determined by the invariants introduced
above: \be 2(\nu^{\mp})^2 =
\Delta_{\sig}\mp\sqrt{\Delta_{\sig}^2-4\,{\rm Det}\,\sig}
\;\label{symeig} . \ee Also the purity $\mu={\rm Tr}\,\varrho^2$ of
a multimode Gaussian state $\varrho$, quantifying its degree of
mixedness, is easily determined in terms of the symplectic
invariants $\,{\rm Det}\,\sig$, as  \cite{paris03} \be
\mu=1/\sqrt{\,{\rm Det}\,\sig}\,. \label{muparis}\ee

Regarding the entanglement of Gaussian states, we recall that the
positivity of the partial transpose is a necessary and sufficient
criterion for two-mode states to be separable (PPT criterion)
\cite{simon00}. The validity of such a criterion has been later
extended to generic Gaussian states of $1\times n$-mode systems
\cite{werner01} and to $(m + n)$-mode Gaussian states with fully
degenerate symplectic spectrum \cite{botero03,giedkeqic03}. For a
bipartite system with Hilbert space ${\cal H}={\cal H}_A\otimes
{\cal H}_B$, made up of two subsystems with Hilbert spaces ${\cal
H}_{A}$ and ${\cal H}_{B}$, the operation of partial transposition
is defined as the transposition of the degrees of freedom associated
to only one of the two subsystems, {\em i.e.~}to the transposition
of only one of the reduced Hilbert spaces, say ${\cal H}_{A}$. Let
us remark that the positivity of the partially transposed operator
$\tilde{\varrho}$ does not depend on which subsystem is transposed
nor on the basis chosen to perform the transposition. Therefore the
positivity of the partial transpose is invariant under local unitary
transformations on the two subsystems. In particular, for two-mode
Gaussian states, the PPT criterion reduces to a simple inequality on
the smallest symplectic eigenvalue $\tilde{\nu}^{-}$ of the
partially transposed CM $\tilde{\sig}$ (partial transposition
amounts to the mirror reflection of one of the four quadratures, see
Ref.~\cite{simon00}). A two-mode Gaussian state is separable ({\em
i.e.~}not entangled) if and only if \be \tilde{\nu}^{-} \ge 1 \; .
\label{symppt} \ee

A proper quantification of the entanglement, easily computable for
two-mode Gaussian states, is provided by the negativity $\N$,
thoroughly discussed and extended in Ref.~\cite{vidal02} to CV systems
(see also Refs.~\cite{zircone,jensth}).
The negativity of a
quantum state $\varrho$ is defined as \be {\cal
N}(\varrho)=\frac{\|\tilde \varrho \|_1-1}{2}\: , \ee where
$\tilde\varrho$ is the partially transposed density matrix and
$\|\hat o\|_1=\,{\rm Tr}|\hat o|$ stands for the trace norm of $\hat
o$. The quantity ${\cal N} (\varrho)$ is equal to
$|\sum_{i}\lambda_{i}|$, the modulus of the sum of the negative
eigenvalues of $\tilde\varrho$, quantifying the extent to which
$\tilde\varrho$ fails to be positive. Strictly related to $\N$ is
the logarithmic negativity $E_{\N}$, defined as $E_{\N}\equiv
\ln\|\tilde{\varrho}\|_{1}$, which constitutes an upper bound to the
{\em distillable entanglement} of the quantum state $\varrho$ and is
related to the entanglement cost under PPT preserving operations
\cite{auden03}. It can be easily shown \cite{adesso03bis} that the
logarithmic negativity of a two-mode Gaussian state is a simple
function of the partially transposed symplectic eigenvalue
$\tilde{\nu}^{-}$ alone: \be E_{\N} = \max[0,-\ln\tilde{\nu}^{-}] \;
, \ee quantifying the extent to which Inequality
(\ref{symppt}) is violated.

Let us recall that the bipartite
entanglement of formation $E_{F}$ \cite{bennett96} of a quantum state $\varrho$,
shared by parties $A$ and $B$, is defined as
\be
E_{F}(\varrho)=\min_{\{p_i,\ket{\psi_i}\}}\sum_i p_i E(\ket{\psi_i}) \; ,
\label{eof}
\ee
where the minimum is taken over all the pure states realizations of $\varrho$:
\[
\varrho=\sum_i p_i \ket{\psi_i}\bra{\psi_i}
\]
and $E(\ket{\psi_i})$ denotes the entropy of entanglement of
the pure state $\ket{\psi_i}$, defined as the von Neumann entropy
of the reduced state obtained by tracing over the variables of one of the
two subsystems:
$$E(\ket{\psi_i})=-\,{\rm Tr_A}[\,{\rm Tr}_B \ket{\psi_i}
\bra{\psi_i}\ln(\,{\rm Tr}_B \ket{\psi_i}\bra{\psi_i})] \, .$$ As
far as symmetric ({\em i.e.~}with ${\rm Det}\,\sig_{11}=\,{\rm
Det}\,\sig_{22}$, with reference to the decomposition of \eq{subma})
two-mode Gaussian states are concerned, the entanglement of formation
$E_{F}$, can be computed \cite{giedkeprl03}. The quantity
$E_{F}$ turns out to be, again, a decreasing function of
$\tilde{\nu}^{-}$: \be E_F = \max\left[ 0,h(\tilde{\nu}^{-}) \right]
\; , \label{eofgau} \ee with
\[
h(x)=\frac{(1+x)^2}{4x}\ln\left(\frac{(1+x)^2}{4x}\right)-
\frac{(1-x)^2}{4x}\ln\left(\frac{(1-x)^2}{4x}\right) \, .
\]
Therefore the entanglement of formation provides, for two-mode
symmetric Gaussian states, a quantification of entanglement fully equivalent
to the one provided by the logarithmic negativity $E_{\N}$.

\section{Standard forms of bisymmetric multimode Gaussian states
\label{sfbisym}}

We shall say that a multimode Gaussian state $\varrho$
is {\it fully symmetric} if it is invariant under the exchange
of any two modes.
In the following, we will consider the fully symmetric $m$-mode and
$n$-mode Gaussian states $\varrho_{\alp^m}$ and $\varrho_{\bet^n}$,
with CMs $\sig_{\alp^m}$ and $\sig_{\bet^n}$.
Due to symmetry, we have that
\begin{equation}\label{fscm}
\sig_{\alp^m}={\left(%
 \begin{array}{cccc}
  \gr\alpha & \gr\varepsilon & \cdots & \gr\varepsilon \\
  \gr\varepsilon & \gr\alpha & \gr\varepsilon & \vdots \\
  \vdots & \gr\varepsilon & \ddots & \gr\varepsilon \\
  \gr\varepsilon & \cdots & \gr\varepsilon & \gr\alpha \\
\end{array}%
\right)}\,, \quad
\sig_{\bet^n}={\left(%
 \begin{array}{cccc}
  \gr\beta & \gr\zeta & \cdots & \gr\zeta \\
  \gr\zeta & \gr\beta & \gr\zeta & \vdots \\
  \vdots & \gr\zeta & \ddots & \gr\zeta \\
  \gr\zeta & \cdots & \gr\zeta & \gr\beta \\
\end{array}%
\right)}\,,
\end{equation}
where $\gr\alpha$, $\gr\varepsilon$, $\gr\beta$ and $\gr\zeta$ are $2\times2$ real
symmetric submatrices (the symmetry of $\gr\varepsilon$ and $\gr\zeta$ stems again
from the symmetry under the exchange of any two modes).
All the properties related to correlations and entropic measures of multimode
Gaussian states are invariant under local, single-mode symplectic operations.
A first preliminary fact, analogous to the standard form reduction of two-mode states,
will thus prove useful. \smallskip

\noindent{\bf Standard form of fully symmetric states.}
{\em Let $\sig_{\beta^n}$ be the CM of a fully symmetric $n$-mode
Gaussian state.
The $2\times 2$ blocks $\bet$ and $\gr\zeta$ of $\sig_{\beta^n}$,
defined by \eq{fscm}, can be brought by means of
local, single-mode symplectic operations $S\in \sy{2}^{\oplus n}$ into the
form $\bet=\,{\rm diag}\,(b,b)$ and
$\gr\zeta=\,{\rm diag}\,(z_1,z_2)$.}\medskip

\noindent {\em Proof.} The blocks $\bet$, being CM's of
reduced single mode Gaussian states, can be turned into their
Williamson standard form by the same symplectic $S_l\in\sy{2}$
acting on each mode. One is then left with the freedom of applying local,
single-mode rotations that leave the blocks $\bet$ invariant.
The same rotation applied to each mode is sufficient to diagonalize
$\gr\zeta$, since such a matrix is symmetric. \hfill$\Box$\smallskip

The coefficients $b$, $z_1$, $z_2$ of the standard form are
determined by the local, single-mode invariant
$\det{\bet}\equiv\mu_{\beta}^{-2}$, and by the symplectic invariants
$\det{\sig_{\beta^2}}\equiv \mu_{\beta^2}^{-2}$ and $\D_2\equiv
\D(\sig_{\beta^2})$. Here $\mu_{\beta}$ ($\mu_{\beta^2}$) is the
marginal purity of the single-mode (two-mode) reduced states, while
$\D_2$ is the remaining invariant of the two-mode reduced states
\cite{adesso03}. This parametrization is provided, in the present
instance, by the following equations \be b=\frac{1}{\mu_{\beta}} \,
, \quad z_1=\frac{\mu_{\beta}}{4} (\epsilon_- - \epsilon_+)\,,\quad
z_2=\frac{\mu_{\beta}}{4} (\epsilon_- +
\epsilon_+)\,,\label{parasim} \ee
\[
{\rm with}\quad
\epsilon_{-}=\sqrt{\D_2^2-\frac{4}{\mu_{\beta^2}^2}}\,,
\]
\[
{\rm and}\quad\epsilon_{+} =
\sqrt{\left(\D_2-\frac{4}{\mu_{\beta}^2}\right)^2-\frac{4}{\mu_{\beta^2}^2}}
\, .
\]
This parametrization has a straightforward interpretation,
because $\mu_{\beta}$ and $\mu_{\beta^2}$
quantify the local mixednesses and $\D_2$ regulates the entanglement
of the two-mode blocks at fixed global and local purities \cite{adesso03}.

Let us next determine and analyse the symplectic
spectrum (symplectic eigenvalues) of $\sig_{\beta^n}$.\smallskip

\noindent{\bf Symplectic degeneracy of fully symmetric states.}
{\em The symplectic spectrum of $\sig_{\beta^n}$ is $n-1$ times
degenerate. The two symplectic eigenvalues of $\sig_{\beta^n}$
$\nu_{\beta}^{-}$ and $\nu_{\beta^n}^{+}$ read
\be
\begin{split}
\nu_{\beta}^{-}&  =    \sqrt{(b-z_1)(b-z_2)} \; ,\\
\nu_{\beta^n}^{+}& =    \sqrt{(b+(n-1)z_1)(b+(n-1)z_2)} \; ,
\end{split}\label{fsspct}
\ee
where $\nu_{\beta}^{-}$ is the $(n-1)$-times degenerate eigenvalue.}\medskip

\noindent {\em Proof.}
We recall that the symplectic eigenvalues of $\sig_{\beta^n}$ are
the absolute values of the eigenvalues of $i\gr{\Omega}\sig_{\beta^n}$.
Since the symplectic form $\gr{\Omega}$ is block diagonal, with
$2\times 2$ blocks $\gr{\omega}$ given by \eq{symform},
the matrix $i\gr{\Omega}\gr{\sigma}$ is
just the matrix $\sig$ with $i\gr{\omega}$
multiplying on the left any $2\times 2$ block.
Let us now consider the set of vectors $\{v_i\}$,
for $i=1,\ldots,n-1$:
\be
v_i=(0,\ldots,0,\underbrace{v^{\sf T}}_{{\rm mode}\,i},
\underbrace{-v^{\sf T}}_{{\rm mode}\,i+1},
0,\ldots0)^{\sf T} \label{vi1}
\ee
where, for covenience, we have introduced the two-dimensional vector
$v=(i\frac{b-z_2}{\nu^-_{\beta}},1)^{\sf T}$.
The $v_i$ are $n-1$ linear independent vectors. One has
\be
i\gr{\Omega}\gr{\sigma}v_i = i(0,\ldots,0,
\underbrace{(\gr{\omega}(\gr{\beta}-\gr{\zeta})v)^{\sf T}}_{{\rm mode}\,i},
\underbrace{-(\gr{\omega}(\gr{\beta}-\gr{\zeta})v)^{\sf T}}_{{\rm mode}\,i+1},
0,\ldots,0)^{\sf T} \, .\label{auto}
\ee
 A straightforward computation
gives
\bea
i\gr{\omega}(\gr{\beta}-\gr{\zeta})v&=&
i\left(\!\!\begin{array}{cc}
0&1\\
-1&0
\end{array}\!\!\right)
\left(\!\!\begin{array}{cc}
b-z_1&0\\
0&b-z_2
\end{array}\!\!\right)
\left(\!\!\begin{array}{c}
i\sqrt{\frac{b-z_2}{b-z_1}}\\
1
\end{array}\!\!\right)\nonumber\\
&=&\nu^{-}_{\beta} \left(\!\!\begin{array}{c}
i\sqrt{\frac{b-z_2}{b-z_1}}\\
1
\end{array}\!\!\right) \,,
\eea
which recasts \eq{auto} into $i\gr{\Omega}\gr{\sigma}v_i=\nu_{\beta}^{-} v_i$, thus
proving that the symplectic eigenvalue $\nu_{\beta}^{-}$
of $\gr{\sigma}$ is $n-1$ times degenerate.
Note that, as one should expect, there exist also $n-1$ eigenvectors
associated to the negative eigenvalue $-\nu_{\beta}^{-}$. To this end,
it suffices to turn $v$ into $(-i\frac{b-z_2}{\nu_{\beta}^{-}},1)^{\sf T}$.

The remaining linearly independent eigenvector of $i\gr{\Omega}\sig_{\beta^n}$
is the vector
$$(w^{\sf T},\ldots,w^{\sf T})^{\sf T}\;,$$
$${\rm with}\quad
w^{\sf T}=(i\sqrt{b+(n-1)z_1},\sqrt{b+(n-1)z_2}) \: .$$ It is
immediate to verify that such a vector is associated to the
eigenvalue $\nu_{\beta^{n}}^{+}$, completing the proof.
\hfill$\Box$\smallskip

The $(n-1)$-times degenerate eigenvalue $\nu_{\beta}^-$
is independent of $n$,
while $\nu_{\beta^n}^{+}$ can be simply expressed as a function of the
single mode purity $\mu_\beta$ and the
symplectic spectrum of the two-mode block with
eigenvalues $\nu_{\beta}^{-}$ and $\nu_{\beta^2}^{+}$:
\begin{equation}\label{fsnupiun}
(\nu_{\beta^n}^{+})^2 = -\frac{n(n-2)}{\mu_\beta^2}+\frac{(n-1)}
2\left( n({\nu^{+}_{\beta^2}})^2+(n-2)(\nu^{-}_{\beta})^2 \right)\,.
\end{equation}
In turn, the two-mode symplectic eigenvalues are
determined by the two-mode invariants by the relation
\be
2(\nu^{\mp}_{\beta})^2 = \Delta_{\beta^2}\mp\sqrt{\Delta_{\beta^2}^2-4/\mu_{\beta^2}^2} \; .
\label{symp2}
\ee The global purity \eq{muparis} of a fully symmetric multimode
Gaussian state is
\begin{equation}\label{fsmu}
\mu_{\beta^n}\equiv\left(\det\sig_{\beta^n}\right)^{-1/2}=
\left((\nu_\beta^-)^{n-1} \nu_{\beta^{n}}^+\right)^{-1}\,,
\end{equation}
and, through \eq{fsnupiun}, can be fully determined in terms
of the one- and two-mode parameters alone.

Obviously, analogous results hold for the $m$-mode CM
$\sig_{\alpha^m}$ of \eq{fscm}, whose $2\times 2$ submatrices can be
brought to the form $\alp=\,{\rm diag}\,(a,a)$ and
$\gr\varepsilon=\,{\rm diag}\,(e_1,e_2)$ and whose $(m-1)$-times
degenerate symplectic spectrum reads \be
\begin{split}
\nu_{\alpha}^{-}&  =    (a-e_1)(a-e_2) \; ,\\
\nu_{\alpha^m}^{+}& =    (a+(m-1)e_1)(a+(m-1)e_2) \; .
\end{split}\label{fsspct2}
\ee

Let us now generalize this analysis to the $(m+n)$-mode Gaussian states with CM $\sig$,
which results from a correlated combination of the fully symmetric blocks
$\sig_{\alp^m}$ and $\sig_{\bet^n}$:
\be
\sig = \left(\begin{array}{cc}
\sig_{\alp^m} & \gr\Gamma\\
\gr\Gamma^{\sf T} & \sig_{\bet^n}
\end{array}\right) \; , \label{fulsim}
\ee where $\gr\Gamma$ is a $2m\times 2n$ real matrix formed by
identical $2\times 2$ blocks $\gr\gamma$.
Clearly, $\gr\Gamma$ is responsible of the correlations existing
between the $m$-mode and the $n$-mode parties. Once again, the
identity of the submatrices $\gr\gamma$ is a consequence of the
local invariance under mode exchange, internal to the $m$-mode
and $n$-mode parties. States of the form of \eq{fulsim} will be
henceforth referred to as {\it bisymmetric}. A significant insight
into bisymmetric multimode Gaussian states can be gained by studying
the symplectic spectrum of $\sig$ and comparing
it to the ones of $\sig_{\alpha^m}$ and
$\sig_{\beta^n}$.\smallskip

\noindent{\bf Symplectic degeneracy of bisymmetric states.}
{\em The symplectic spectrum of the
CM $\sig$ \eq{fulsim} of a bisymmetric $(m+n)$-mode Gaussian
state includes two degenerate eigenvalues,
with multiplicities $m-1$ and $n-1$. Such eigenvalues coincide,
respectively, with the degenerate eigenvalue $\nu_{\alpha}^{-}$ of the
reduced CM $\sig_{\alpha^m}$ and the degenerate eigenvalue
$\nu_{\beta}^{-}$ of the reduced CM $\sig_{\beta^n}$.}\medskip

\noindent{\em Proof.} One can proceed constructively, in
analogy with the proof of the previous proposition. Let us consider
the standard forms of the blocks $\sig_{\alpha^m}$ and $\sig_{\beta^n}$,
while keeping the $2\times 2$ submatrices $\gr\gamma$ in arbitrary,
generally nonsymmetric, form. Let us next focus on the block
$\sig_{\beta^n}$ and define the vectors $\bar{v}_i$
by \be \bar{v}_i=(0,\ldots,0,v_i^{\sf T})^{\sf T} \; . \label{vi}
\ee They are the vectors obtained from the vectors $v_i$'s of
\eq{vi1} by appending to them $2m$ null entries on the
left. Because of the identity of the blocks $\gr\gamma$, their
contributions to the secular equation cancel out and it is
straightforward to verify that the vectors $\bar{v}_i$'s
are $n-1$ eigenvectors of $i\gr{\Omega}\sig$ with eigenvalue
$\nu_{\beta}^{-}$. The same argument holds considering the submatrix
$\sig_{\alpha^m}$, thus completing the proof. \hfill$\Box$\smallskip

Equipped with these results, we are now in a position to
determine the bipartite entanglement of bisymmetric
multimode Gaussian states and prove that it can always
be unitarily {\it localized} or {\it concentrated}. \smallskip

\noindent{\bf Unitary localization of the entanglement of
bisymmetric states.}
{\em The bisymmetric $(m+n)$-mode Gaussian state with CM $\sig$
\eq{fulsim} can be brought, by means of a local unitary operation,
with respect to the $m\times n$-mode bipartition with reduced CMs
$\sig_{\alpha^m}$ and $\sig_{\beta^n}$, to a tensor product of
single-mode uncorrelated states and of a two-mode
Gaussian state.}\medskip

\noindent{\em Proof.} Let us focus on the $n$-mode block
$\sig_{\beta^{n}}$. The vectors $\bar{v}_i$ of \eq{vi},
with the first $2m$ entries equal to $0$, are, by construction,
simultaneous eigenvectors of $i\gr{\Omega}\sig_{\beta^n}$ and
$i\gr{\Omega}\sig$, with the same (degenerate) eigenvalue. This
fact suggests that the phase-space modes corresponding to such
eigenvectors are the same for $\gr{\sigma}$ and for
$\gr{\sigma}_{\beta^{n}}$. Then,
bringing by means of a local symplectic operation
the CM $\gr{\sigma}_{\beta^{n}}$ in Williamson form,
any $(2n-2)\times(2n-2)$ submatrix of $\gr{\sigma}$ will
be diagonalized because the normal modes are common to the
global and local CMs. In other words, no correlations between
the $m$-mode party with reduced CM $\sig_{\alpha^m}$ and such modes
will be left: all the correlations between the $m$-mode and $n$-mode parties
will be concentrated in the two conjugate quadratures of a single mode
of the $n$-mode block. Going through the same argument for the $m-$mode
block with CM $\sig_{\alpha^m}$ would prove the proposition and show
that the whole entanglement between the two multimode blocks
can always be concentrated in only two modes, one for each of the
two multimode parties.

To prove this property we proceed first by investigating
the relationship between the transformations
which diagonalize $i\gr{\Omega}\gr{\sigma}$ and
the symplectic operations that bring $\gr{\sigma}$
in Williamson normal form $\gr{\nu}$ \cite{arnold}.
The problem
one is immediately faced with is that these
transformations are not unique because
the normal form associated to $\gr{\sigma}$ is invariant
under local rotations (this local freedom is always present
in the selection of normal modes) {\it and}, due to
degeneracy, also under global symplectic rotations
of the modes associated to the degenerate eigenvalue $\nu^{-}_{\beta}$.
Thus there is an ambiguity in selecting the eigenvectors of
$i\gr{\Omega}\gr{\sigma}$ and therefore in determining the transformation
that diagonalizes it.
Moreover, if $\{w_i\}$ is a set of $2(m+n)$ column-vectors
normalized eigenvectors of $i\gr{\Omega}\gr{\sigma}$, then any
matrix $T$ of the form
\be
T=\left(\xi_1 w_1,\cdots,\xi_k w_k\right),
\ee
diagonalizes $i\gr{\Omega}\gr{\sigma}$: $T^{-1}(i\gr{\Omega}\gr{\sigma})T=D$
(with the $\xi_i$'s arbitrary complex coefficients).
However, we can proceed by observing that
the $2\times 2$ matrix $i\gr{\omega}$ is
diagonalized by the unitary transformation $\bar{U}$, with
\[
\bar{U}=\frac{1}{\sqrt{2}}\left(\begin{array}{cc}
i&-i\\
1&1
\end{array}\right) \; ,
\]
so that $\bar{U}^{\dag}ia\gr{\omega}\bar{U}=\,{\rm diag}\,(a,-a)$
(where $a$ is any complex number). We can then define the
matrix $U=\bar{U}^{\oplus (m+n)}$, which is local in the sense
that it is block diagonal and acts on each mode separately, such
that for any normal form $\gr{\nu}$
\be
U^{-1}i\gr{\Omega}\gr{\nu}U = D \, ,
\ee
where $D=T^{-1}i\gr{\Omega}\gr{\sigma}T$ is a diagonal matrix
with entries $\{\mp\nu_i\}$ (in terms of the symplectic
eigenvalues). Let us next denote by $S$ one of the symplectic
transformations that bring $\gr{\sigma}$ in normal form: $S^{\sf
T}\gr{\sigma}S=\gr{\nu}$. It is then easy to see that
\be
\begin{split}
D&=T^{-1}(i\gr{\Omega}\gr{\sigma})T=U^{-1}(i\gr{\Omega}\gr{\nu})U\\
&= U^{-1}(i\gr{\Omega} S^{\sf T} \gr{\sigma} S)U =
U^{-1}S^{-1}(i\gr{\Omega}\gr{\sigma})SU \; , \label{sympli}
\end{split}
\ee
and therefore
\be
S=TU^{-1}=TU^{\dag} \; , \label{loca}
\ee
where in \eq{sympli} we have exploited the fundamental property
of symplectic transformations:
$S^{-1\sf T}\gr{\Omega}S^{-1}=\gr{\Omega}$.
\eq{loca} shows that there must exist {\em some} symplectic
transformation that diagonalizes $i\gr{\Omega}\gr{\sigma}$
and satisfies the further condition given by \eq{loca}.
In fact, it is obvious that not every $T$ diagonalizing
$i\gr{\Omega}\gr{\sigma}$ is a symplectic transformation when
multiplied on the right by $U^{\dag}$. Viceversa, if this last
condition holds, the symplectic operation that brings
$\gr{\sigma}$ in normal form is given by \eq{loca}. The modes that
diagonalize the quadratic form $\gr{\sigma}$ in phase space
can be reconstructed in terms of $S$: since they are
linear combinations of the original modes and
$S^{\sf T}\gr{\sigma}S$ is diagonal, they can be expressed
by real column vectors identified by the columns of $S$.

We can now go back to our original problem: leaving aside the involved
task of exactly determining which choice of the eigenvectors of
$i\gr{\Omega}\gr{\sigma}$ leads to a symplectic transformation of the form
\eq{loca}, we are anyway assured that in the subspace associated to the
eigenvalues $\mp\nu_{\beta}^-$ such eigenvectors must be
linear combinations of the $\bar{v}_i$'s defined in \eq{vi}
and their counterparts associated to the
eigenvalue $-\nu_{\beta}^-$
(with their first $2m$ entries, related to the
$m$-mode party, set equal to $0$).
Therefore the transformation $T$ reads, in general,
\be
T=\left(\begin{array}{cccccc}
T_{1,1}&\cdots&T_{1,m}&\gr{0}&\cdots&\gr{0}\\
\vdots&\ddots&\vdots&\vdots&\ddots&\vdots\\
T_{m,1}&\cdots&T_{m,m}&\gr{0}&\cdots&\gr{0}\\
T_{m+1,1}&\cdots&T_{m+1,m}&T_{m+1,m+1}&\cdots&T_{m+1,m+n}\\
\vdots&\ddots&\vdots&\vdots&\ddots&\vdots\\
T_{m+n,1}&\cdots&T_{m+n,m}&T_{m+n,m+1}&\cdots&T_{m+n,m+n}
\end{array}\right) ,
\ee
where $\gr{0}$ stands for $2\times 2$ null matrices and
$T_{i,j}$ are $2\times 2$ blocks, whose exact form is unessential
to our aims.
Exploiting \eq{loca}, for the last $2(n-1)$
columns of $S$ we obtain, in terms of $2\times 2$
matrices,
\be
(\underbrace{\gr{0},\ldots,\gr{0}}_{{\rm first}\;m\;{\rm  modes}},
\bar{U}^{*}T_{2,i}^{\sf T},
\ldots,\bar{U}^{*}T_{n,i}^{\sf T})^{\sf T} \; . \label{nmodes}
\ee
Due to the presence of the first $m$ null entries,
the $n-1$ modes determined by \eq{nmodes}
are normal modes of both the {\em global} CM $\sig$ and the
{\em local} CM $\sig_{\beta^n}$. An analogous proof, going along
the same lines of reasoning, holds for the reduced CM
$\sig_{\alpha^m}$: it can be reduced to a local normal form
that shares $m-1$ normal modes with the global CM $\sig$.
These results imply that the form in which all the correlations
between the two parties are shared only by a single mode of the $n$-mode
party and by a single mode of the $m$-mode party can be obtained by means
of local symplectic (unitary) operations, namely by the symplectic
operations bringing the block $\gr{\sigma}_{\beta^{n}}$ and the block
$\gr{\sigma}_{\alpha^{m}}$ in Williamson form.

For ease of the reader and sake of pictorial clarity, we can
supplement the proof by explicitly writing down the different forms
of the CM $\sig$ at each step; such matrix representations allow an immediate
visualization of the process of unitary concentration of the entanglement
between a single pair of modes, one for each multimode party.
The CM $\sig$ of a bisymmetric $(m+n)$-mode Gaussian state
reads (see \eq{fulsim})
\be
\gr{\sigma}=
\left(\begin{array}{cccccccc}
\gr{\alpha}&\gr{\varepsilon}&\ldots&\gr{\varepsilon}&
\gr{\gamma}&\cdots&\cdots&\gr{\gamma}\\
\gr{\varepsilon}&\ddots&\gr{\varepsilon}&\vdots&
\vdots&\ddots&&\vdots\\
\vdots&\gr{\varepsilon}&\ddots&\gr{\varepsilon}&
\vdots&&\ddots&\vdots\\
\gr{\varepsilon}&\cdots&\gr{\varepsilon}&\gr{\alpha}&
\gr{\gamma}&\cdots&\cdots&\gr{\gamma}\\
\gr{\gamma}^{\sf T}&\cdots&\cdots&\gr{\gamma}^{\sf T}&
\gr{\beta}&\gr{\zeta}&\ldots&\gr{\zeta}\\
\vdots&\ddots&&\vdots&
\gr{\zeta}&\ddots&\gr{\zeta}&\vdots\\
\vdots&&\ddots&\vdots&
\vdots&\gr{\zeta}&\ddots&\gr{\zeta}\\
\gr{\gamma}^{\sf T}&\cdots&\cdots&\gr{\gamma}^{\sf T}&
\gr{\zeta}&\cdots&\gr{\zeta}&\gr{\beta}
\end{array}\right) \, . \label{totala}
\ee
According to what we have just shown, reducing to normal form
the block $\gr{\sigma}_{\beta^{n}}$ brings the global CM
$\sig$ in the form
CM $\gr{\sigma'}$
\[\gr{\sigma'}=
\left(\begin{array}{cccccccc}
\gr{\alpha}&\gr{\varepsilon}&\cdots&\gr{\varepsilon}&
\gr{\gamma'}&\gr{0}&\cdots&\gr{0}\\
\gr{\varepsilon}&\ddots&\gr{\varepsilon}&\vdots&
\vdots&\vdots&\ddots&\vdots\\
\vdots&\gr{\varepsilon}&\ddots&\gr{\varepsilon}&
\vdots&\vdots&\ddots&\vdots\\
\gr{\varepsilon}&\cdots&\gr{\varepsilon}&\gr{\alpha}&
\gr{\gamma'}&\gr{0}&\cdots&\gr{0}\\
\gr{\gamma'}^{\sf T}&\cdots&\cdots&\gr{\gamma'}^{\sf T}&
\gr{\nu}_{\beta^n}^+&\gr{0}&\cdots&\gr{0}\\
\gr{0}&\cdots&\cdots&\gr{0}&
\gr{0}&\gr{\nu}_{\beta}^-&\gr{0}&\vdots\\
\vdots&\ddots&\ddots&\vdots&
\vdots&\gr{0}&\ddots&\gr{0}\\
\gr{0}&\cdots&\cdots&\gr{0}&
\gr{0}&\cdots&\gr{0}&\gr{\nu}_{\beta}^-
\end{array}\right) ,
\]
where the $2\times 2$ blocks $\gr{\nu}_{\beta^n}^{+}=
\nu_{\beta^n}^{+}{\mathbbm 1}_2$ and $\gr{\nu}_{\beta}^-=
\nu_{\beta}^{-}{\mathbbm 1}_2$ are
the Williamson normal blocks associated to the two
symplectic eigenvalues of $\gr{\sigma}_{\beta^{n}}$.
The identity of the submatrices $\gr{\gamma'}$ is due to the
invariance under permutation of the first $m$ modes,
which are left unaffected.
The subsequent symplectic diagonalization of $\gr{\sigma}_{\alpha^{m}}$
puts the global CM $\sig$ in the following form (notice that the first,
$m+1$-mode reduced CM is again a matrix of the same form of $\sig$,
with $n=1$):
\be
\gr{\sigma''}=
\left(\begin{array}{cccccccc}
\gr{\nu}_{\alpha}^-&\gr{0}&\cdots&\gr{0}&
\gr{0}&\gr{0}&\cdots&\gr{0}\\
\gr{0}&\ddots&\gr{0}&\vdots&
\vdots&\vdots&\ddots&\vdots\\
\vdots&\gr{0}&\gr{\nu}_{\alpha}^-&\gr{0}&
\gr{0}&\vdots&\ddots&\vdots\vspace*{.25cm}\\
\gr{0}&\cdots&\gr{0}&\gr{\nu}_{\alpha^m}^+&
\gr{\gamma''}&\gr{0}&\cdots&\gr{0}\vspace*{.25cm}\\
\gr{0}&\cdots&\gr{0}&\gr{\gamma''}^{\sf T}&
\gr{\nu}_{\beta^n}^+&\gr{0}&\cdots&\gr{0}\\
\gr{0}&\cdots&\cdots&\gr{0}&
\gr{0}&\gr{\nu}_{\beta}^-&\gr{0}&\vdots\\
\vdots&\ddots&\ddots&\vdots&
\vdots&\gr{0}&\ddots&\gr{0}\\
\gr{0}&\cdots&\cdots&\gr{0}&
\gr{0}&\cdots&\gr{0}&\gr{\nu}_{\beta}^-
\end{array}\right) , \label{final}
\ee with $\gr{\nu}_{\alpha^m}^{+}= {\nu}_{\alpha^m}^{+} {\mathbbm
1}_2$ and $\gr{\nu}_{\alpha}^{-}= {\nu}_{\alpha}^{-} {\mathbbm
1}_2$. \eq{final} shows explicitly that the state with CM
$\gr{\sigma''}$, obtained from the original state with CM
$\sig$ by exploiting local unitary operations, is the tensor product
of $m+n-2$ uncorrelated single-mode states and of a correlated
two-mode Gaussian state. The proof is therefore complete, and
shows that the amount of entanglement (quantum correlations) present
in any bisymmetric multimode Gaussian state can be localized
(concentrated) in a two-mode Gaussian state ({\it i.e.}
shared only by a single pair of modes), via local unitary operations.
These results and their consequences will be discussed in detail in
the following sections.  \hfill$\Box$\smallskip

\section{Block entanglement of multimode Gaussian states \label{block}}

In the previous section,
the study of the multimode CM $\sig$ of \eq{totala}
has been reduced to a two-mode problem
by means of local unitary operations.
This finding allows for an exhaustive analysis of the
bipartite entanglement between the $m$- and $n$-mode blocks
of a multimode Gaussian state, resorting to the
powerful results available for two-mode Gaussian states.
For any multimode Gaussian state with CM $\sig$,
let us define the associated {\it equivalent} two-mode
Gaussian state $\varrho_{eq}$,
with CM $\sig_{eq}$ given by
\be
\sig_{eq}= \left(\begin{array}{cc}
\gr{\nu}_{\alpha^m}^{+}&\gr{\gamma''}\\
\gr{\gamma''}^{\sf T}& \gr{\nu}_{\beta^n}^{+}
\end{array}\right) \; ,
\ee where the $2\times 2$ blocks have been implicitly defined in the
CM (\ref{final}). As already mentioned, the entanglement of the bisymmetric
state with CM $\sig$, originally shared among all the $m+n$ modes,
can be {\em completely} concentrated by local unitary (symplectic)
operations on a single pair of modes in the state with CM $\sig_{eq}$.
Such an entanglement is, in this sense, localizable. Obviously, this kind
of localization of entanglement by local unitaries is conceptually
very different from the localization of entanglement by local measurements
first discussed by Verstraete, Popp, and Cirac for qubit systems \cite{vpclocal}.
We now move on to describe some consequences of this result.

A first qualificative remark is in order. It is known that the PPT
criterion is necessary and sufficient for the separability of
Gaussian states of $1 \times 1$-mode and  $1 \times n$-mode
bipartitions. In view of the invariance of such a criterion under
local unitary transformations, which can be appreciated by the
definition of partial transpose at the Hilbert space level, and
considering the results proved in the previous section, it is
immediate to verify that the following property holds:\medskip

\noindent{\bf PPT criterion for bisymmetric multimode Gaussian states.} {\em
For generic $m \times n$-mode bipartitions, the positivity of the partial
transpose (PPT) is a necessary and sufficient condition for the
separability of bisymmetric $m+n$-mode Gaussian states.}
\smallskip

This statement is a first important generalization to $m \times n$
bipartitions of the result proved by Werner and Wolf for the case of
$1 \times n$ bipartitions \cite{werner01}. In particular, it implies
that no bisymmetric bound entangled Gaussian states may exist
\cite{werner01,giedkeqic01} and all the $m \times n$ block
entanglement of such states is distillable. Moreover, it justifies
the use of the negativity and the logarithmic negativity as measures
of entanglement for these multimode Gaussian states.

As for the quantification of the entanglement,
exploiting some recent results on two-mode Gaussian
states \cite{adesso03,adesso03bis} we can
select the relevant quantities that, by determining the
correlation properties of the two-mode Gaussian state with
CM $\sig_{eq}$, also determine the entanglement and correlations
of the multimode Gaussian state with CM $\sig$.
These quantities are, clearly, the equivalent marginal purities $\mu_{\alpha eq}$
and $\mu_{\beta eq}$, the global purity $\mu_{eq}$ and the
equivalent two-mode invariant $\D_{eq}$.
Let us remind that, by exploiting Eqs.~(\ref{fsspct}), (\ref{fsspct2})
and (\ref{parasim}), the symplectic spectra of the CMs
$\gr{\sigma}_{\alpha^{m}}$ and $\sig_{\beta^{n}}$
may be recovered by means of the local two-mode
invariants $\mu_{\beta}$, $\mu_{\alpha}$, $\mu_{\beta^2}$,
$\mu_{\alpha^2}$, $\D_{\beta^2}$ and $\D_{\alpha^2}$.
The quantities $\mu_{\alpha eq}$ and $\mu_{\beta eq}$
are easily determined in terms of local invariants alone:
\be
\mu_{\alpha eq}=1/\nu_{\alpha^m}^{+} \; \quad
\mu_{\beta eq}=1/\nu_{\beta^n}^{+} \; . \label{localone}
\ee
On the other hand, the determination of $\mu_{eq}$ and
$\D_{eq}$ require the additional knowledge of two global
symplectic invariants of the CM $\sig$; this should be expected,
because they are susceptible of quantifying the correlations
between the two parties. The natural choices for the global invariants
are the global purity $\mu=1/\sqrt{\det{\sig}}$
and the invariant $\D$, given by
\bea
\D&=&m\det{\alp}+m(m-1)\det{\gr{\varepsilon}}+
n\det{\bet}\nonumber\\
&&+n(n-1)\det{\gr{\zeta}}+2mn\det{\gr{\gamma}}\nonumber\, .
\eea
One has
\bea
\mu_{eq}&=&(\nu_{\alpha}^{-})^{m-1}(\nu_{\beta}^{-})^{n-1}\mu \; ,\label{mueq}\\
\D_{eq}&=& \D-(m-1)(\nu_{\alpha}^{-})^{2}-(n-1)(\nu_{\beta}^{-})^{2} \; .
\eea

The entanglement, quantified by the logarithmic negativity, and the
mutual information between the $m$-mode and the $n$-mode subsystems
can thus be easily determined, as it is the case for two-mode
states. In particular, the smallest symplectic eigenvalue $\tilde{\nu}_{eq}$
of the matrix $\tilde{\sig}_{eq}$, derived from $\sig_{eq}$ by partial
transposition, fully quantifies the entanglement
between the $m$-mode and $n$-mode partitions.
Recalling the results known for two-mode states \cite{adesso03,adesso03bis},
the quantity $\tilde{\nu}_{eq}$ reads
\bea
2\tilde{\nu}_{eq}^2
&=&\tilde{\D}_{eq}-\sqrt{\tilde{\D}_{eq}^2-
\frac{4}{\mu_{eq}^2}}\, ,\nonumber\\
{\rm with} \quad \tilde{\D}_{eq}&=& \frac{2}{\mu_{\alpha
eq}^2}+\frac{2}{\mu_{\beta eq}^2}- \D_{eq} \, . \nonumber
\eea
The logarithmic negativity $E_{\N}^{\alpha^m|\beta^n}$ measuring the
bipartite entanglement between the $m$-mode and $n$-mode subsystems
is then
\be
E_{\N}^{\alpha^m|\beta^n}= \max\left[-\ln\tilde{\nu}_{eq},0\right]
\; .
\ee

In the case $\nu_{\alpha^m}^{+}=\nu_{\beta^n}^{+}$, corresponding to
the condition \be (a+(m-1)e_1)(a+(m-1)e_2) =
(b+(n-1)z_1)(b+(n-1)z_2) \; , \label{possi} \ee the equivalent
two-mode state is symmetric and we can determine also the entanglement
of formation, using \eq{eofgau}. Let us note that the
possibility of exactly determining the entanglement of formation of
a multimode Gaussian state of a $m\times n$-mode bipartition is a rather
remarkable consequence, even under the symmetry constraints obeyed
by the CM $\sig$. Another relevant fact to point out is that, since both
the logarithmic negativity and the entanglement of formation are
decreasing functions of the quantity $\tilde{\nu}_{eq}$, the two
measures induce the same entanglement hierarchy on such a subset of
equivalently symmetric states ({\em i.e.}~states whose equivalent
two--mode CM $\sig_{eq}$ is symmetric).

From \eq{mueq} it follows that, if the $(m+n)$-mode bisymmetric
state is pure ($\mu=\nu_{\alpha^m}^-=\nu_{\beta^n}^-=1$), then the
equivalent two-mode state is pure as well ($\mu_{eq}=1$) and,
up to local symplectic operations, it is a two-mode
squeezed vacuum. Therefore {\em any pure bisymmetric multimode Gaussian
state is equivalent, under local unitary (symplectic) operations,
to a tensor product of a pure two-mode squeezed vacuum and of
$m+n-2$ uncorrelated vacua}.

More generally, if both the reduced $m$-mode and $n$-mode CMs
$\sig_{\alpha^m}$ and $\sig_{\beta^m}$ of a bisymmetric, mixed
multimode Gaussian state $\sig$ of the form \eq{fulsim} correspond
to Gaussian mixed states of partial minimum uncertainty, {\em i.e.}~if
$\nu_{\alpha^m}^-=\nu_{\beta^n}^-=1$, then \eq{mueq} implies
$\mu_{eq}=\mu$. Therefore, the equivalent two-mode state has the same
entanglement and the same degree of mixedness of the original
multimode state. In all other cases of bisymmetric multimode states
one has that $\mu_{eq} > \mu$ and the process of localization
produces a two--mode state with higher purity than the original
multimode state. In this specific sense, we see that the process of
localization implies a process of purification as well. We can
understand this key point observing that the entanglement is
localized by performing local unitary transformation which are
reversible by definition. Then, in principle, by only using passive
and active linear optics elements such as beam splitters, phase shifters
and squeezers \cite{vloock03}, one can implement a reversible
machine that, from mixed, bisymmetric multimode states with strong
quantum correlations between all the modes (and consequently between
the $m$-mode and the $n$-mode partial blocks) but weak couplewise
entanglement, is able to extract a highly pure, highly entangled
two-mode state (with no entanglement lost, all the $m \times n$
entanglement can be localized). If needed, the same machine would
be able, starting from a two-mode squeezed state and a collection of
uncorrelated thermal or squeezed states, to distribute the two--mode
entanglement between all modes, converting the two-mode into
multimode, multipartite quantum correlations, again with no loss of
entanglement. The bipartite or multipartite entanglement can then be
used on demand, the first for instance in a CV quantum teleportation
protocol, the latter to secure quantum key distribution or to
perform multimode entanglement swapping.

\section{Quantitative localization of the block entanglement \label{examp}}

In this section we will explicitly compute the block entanglement
({\em i.e.~}the entanglement between different blocks of modes) for
some instances of multimode Gaussian states. We will study its scaling
behavior as a function of the number of modes and explore in deeper
detail the localizability of the multimode entanglement. We focus our
attention on fully symmetric $2n$-mode Gaussian states described by
a $2n \times 2n$ CM $\sig_{\beta^{2n}}$ given by \eq{fscm}.
These states are trivially bisymmetric under any bipartition of the modes,
so that their block entanglement is always localizable by means of local symplectic
operations. Let us recall that concerning the covariances in normal
forms of fully symmetric states (see Sec.~\ref{sfbisym}), pure
states are characterized by
\begin{eqnarray}
z_{i} & = & \left[ 1+b^2(2n-2)-(2n-1) - (-1)^{i} \right. \nonumber \\
& \times & \left. \sqrt{(b^{2} -1)((2 b n)^2-(2n-2)^2)}\right]
/\left[ 2b(2n-1)\right] \; , \nonumber \\
&&
\end{eqnarray}
and belong to the class of CV GHZ--type states discussed in Refs.~
\cite{vloock03,adesso04}. These multipartite entangled states are
generated as the outputs of the application of a sequence of $2n-1$
beam splitters to $2n$ single--mode squeezed inputs \cite{vloock03}.
In the limit of infinite squeezing,
these states reduce to the simultaneous eigenstates of the relative
positions and the total momentum, which define the proper GHZ states
of CV systems \cite{vloock03}.
The CM $\sig^{p}_{\beta^{2n}}$ of this class of pure states, for a
given number of modes, depends only on the parameter $b\equiv
1/\mu_\beta \ge 1$, which is an increasing function of the
single-mode squeezing. Correlations between the modes are induced
according to the above expression for the covariances $z_{i}$.
Exploiting our previous analysis, we can compute the entanglement
between a block of $k$ modes and the remaining $2n-k$ modes, both
for pure states (in this case the block entanglement is simply
the Von Neumann entropy of each of the reduced blocks) and,
remarkably, also for mixed states.

\begin{figure}[t!]
\includegraphics[width=7.5cm]{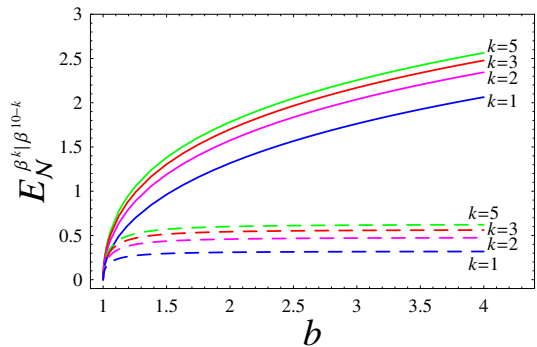}
\caption{(color online). Hierarchy of block entanglements of fully
symmetric $2n$-mode Gaussian states of $k \times (2n-k)$
bipartitions ($n=10$) as a function of the single-mode squeezing
$b$. The block entanglements are depicted both for pure states
(solid lines) and for mixed states obtained from fully symmetric
$(2n+4)$-mode pure Gaussian states by tracing out $4$ modes (dashed
lines). All the quantities plotted are dimensionless.}
\label{fiscalb}
\end{figure}

We can in fact consider a generic $2n$-mode fully symmetric mixed state
with CM $\sig_{\beta^{2n}}^{p\backslash q}$, obtained from a pure
fully symmetric $(2n+q)$-mode state by tracing out $q$ modes. For
any $q$, for any dimension $k$ of the block ($k \leq n$), and for any
non zero squeezing ({\em i.e.~}for $b>1$) one has that $\tilde{\nu}_k<1$,
meaning that the state exhibits genuine multipartite entanglement,
as first remarked in Ref. \cite{vloock03} for pure states: each
$k$-mode party is entangled with the remaining $(2n-k)$-mode
block. Furthermore, the genuine multipartite nature of the
entanglement can be precisely quantified by observing that
$E_\N^{\beta^k\vert\beta^{2n-k}}$ is an increasing function of the
integer $k \le n$, as shown in Fig.~\ref{fiscalb}. Moreover, we
note that the multimode entanglement of mixed states remains finite
also in the limit of infinite squeezing, while the multimode entanglement
of pure states diverges with respect to any bipartition, as shown
in Fig.~\ref{fiscalb}.

In fully symmetric Gaussian states, the block entanglement is
localizable with respect to any $k \times (2n-k)$ bipartition. Since
in this instance {\em all} the entanglement can be concentrated on
a single pair of modes, after the partition has been decided, no strategy
could grant a better yield than the local symplectic operations
bringing the reduced CMs in Williamson form (because of the
monotonicity of the entanglement under general LOCC). However, the
amount of block entanglement, which is the amount of concentrated
two--mode entanglement after unitary localization hase taken place,
actually depends on the choice of a particular $k \times (2n-k)$ bipartition,
giving rise to a hierarchy of localizable entanglements.

Let us suppose that a given Gaussian multimode state (say, for simplicity,
a fully symmetric state) is available and its entanglement is
meant to serve as a resource for a given protocol. Let us further
suppose that the protocol is optimally implemented if the
entanglement is concentrated between only two modes of the global
systems, as it is the case, {\em e.g.}, in a CV teleportation
protocol between two single-mode parties.
Which choice of the bipartition between the modes allows
for the best entanglement concentration by a succession of local
unitary operations?
In this framework, for an even number of modes,
the worst localization strategy consists
in assigning $k=1$ mode at one party and $2n-1$ modes to the other.
Conversely, the best option for localization is an equal $k=n$
splitting of the $2n$ modes between the two parties.  The
logarithmic negativity $E_\N^{\beta^n\vert\beta^{n}}$, concentrated
into two modes by local operations, represents the optimal
localizable entanglement (OLE) of the state $\sig_{\beta^{2n}}$,
where ``optimal'' refers to the choice of the bipartition. Clearly,
the OLE of a state with $2n+1$ modes is given by
$E_\N^{\beta^{n+1}\vert\beta^{n}}$. These
results may be applied to arbitrary, pure or mixed, fully
symmetric Gaussian states.

\begin{figure}[t!]
\includegraphics[width=7.8cm]{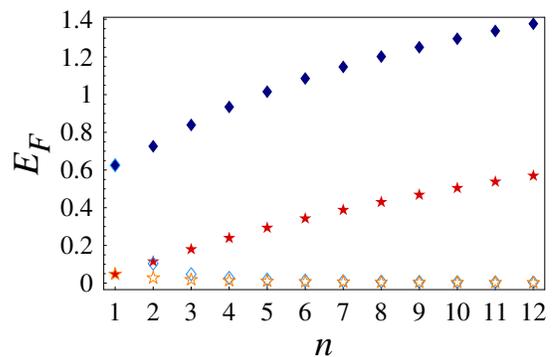}
\caption{(color online). Scaling, with half the number of modes, of
the entanglement of formation in two families of fully symmetric
$2n$-mode Gaussian states. Diamonds denote pure states, while mixed
states (denoted by stars) are obtained from $(2n+4)$-mode pure
states by tracing out $4$ modes. For each class of states, two sets
of points are plotted, one referring to $n \times n$ entanglement
(filled symbols), and the other to $1 \times 1$ entanglement (empty
symbols). Notice how the $n \times n$ entanglement, equal to the
optimal localizable entanglement (OLE) and estimator of genuine
multipartite quantum correlations among all the $2n$ modes,
increases at the detriment of the bipartite $1\times1$ entanglement
between any pair of modes. The single-mode squeezing parameter is
fixed at $b=1.5$. All the quantities plotted are dimensionless.}
\label{fisnef}
\end{figure}

We now turn to the study of the scaling behavior with $n$ of the OLE
of $2n$-mode states, to understand how the number of local
cooperating parties can improve the maximal entanglement that can be
shared between two parties. For generic (mixed) fully symmetric
$2n$-mode states of $n \times n$ bipartitions, the OLE can be
quantified also by the entanglement of formation $E_F$, as the
equivalent two-mode state is symmetric. It is then useful to
compare, as a function of $n$, the $1 \times 1$ entanglement of
formation between a pair of modes (all pairs are equivalent due to
the global symmetry of the state) before the localization, and the
$n \times n$ entanglement of formation, which is equal to the
optimal entanglement concentrated in a specific pair of modes after
performing the local unitary operations. The results of this study
are shown in Fig. \ref{fisnef}. The two quantities are plotted at
fixed squeezing $b$ as a function of $n$ both for a pure $2n$-mode
state with CM $\sig_{\beta^{2n}}^p$ and a mixed $2n$-mode state with
CM $\sig_{\beta^{2n}}^{p\backslash4}$. As the number of modes
increases, any pair of modes becomes steadily less entangled, but
the total multimode entanglement of the state grows and, as a
consequence, the OLE increases with $n$. In the limit $n \rightarrow
\infty$, the $n \times n$ entanglement diverges while the $1\times1$
one vanishes. This holds both for pure and mixed states, although
the global degree of mixedness produces the typical behavior that
tends to reduce the total entanglement of the state.

\section{Concluding Remarks \label{conclu}}

We have shown that bisymmetric multimode Gaussian states (pure or
mixed) can be reduced, by local symplectic operations, to the tensor
product of a correlated two-mode Gaussian state and of uncorrelated
thermal states (the latter being obviously irrelevant as far as the
correlation properties of the multimode Gaussian state are
concerned). As a consequence, {\it all} the entanglement of
bisymmetric multimode Gaussian states of arbitrary $m \times n$
bipartitions is {\em unitarily localizable} in a single (arbitrary)
pair of modes shared by the two parties. Such a useful reduction to
two-mode Gaussian states is somehow similar to the
one holding for states with fully degenerate symplectic spectra
\cite{botero03,giedkeqic03}, encompassing the relevant instance of
pure states, for which all the symplectic eigenvalues are equal to
$1$. The present result allows to extend the PPT criterion as a
necessary and sufficient condition for separability for all
bisymmetric multimode Gaussian states of arbitrary $m \times n$
bipartitions, and to quantify their entanglement.

Notice that, in the general bisymmetric instance addressed in this
work, the possibility of performing a two-mode reduction is
crucially partition-dependent. However, as we have explicitly shown,
in the case of fully symmetric states all the possible bipartitions
can be analysed and compared, yielding remarkable insight into the
structure of the multimode block entanglement of Gaussian states.
This leads finally to the determination of the maximum, or optimal
localizable entanglement  that can be concentrated on a single pair
of modes.

It is important to notice that the multipartite entanglement
in the considered class of multimode Gaussian states can be produced
and detected \cite{vloock00,vloock03}, and also, by virtue of the
present analysis, reversibly localized by all-optical means.
Moreover, the multipartite entanglement
allows for a reliable ({\em i.e.}~with fidelity ${\cal F}>{\cal F}_c$,
where ${\cal F}_c=1/2$ is the classical threshold) quantum teleportation
between any two parties with the assistance of the remaining others
\cite{vloock00}. This quantum teleportation network has been recently
demonstrated experimentally with the use of fully symmetric three-mode
Gaussian states \cite{naturusawa}.

\acknowledgments{We thank INFM, INFN, and MIUR under national
project PRIN-COFIN 2002 for financial support.}


\begin{thebibliography}{99}
\bibitem{absolute} Here $|A|$ stands for the absolute value of the diagonalizable
matrix $A$, to be understood in the usual operatorial sense: if $A=T^{-1}DT$,
where $D$ is diagonal and $T$ is a diagonalizing operator, then $|A|=T^{-1}|D|T$,
where $|D|$ is the diagonal matrix whose entries are the absolute
values of the eigenvalues of $A$.
\end{thebibliography}
\end{document}